\documentclass[osajnl,preprint,showpacs,superscriptaddress,12pt]{revtex4-1} %% use 12pt for preprint option
\usepackage{graphicx}

\usepackage{amsmath,amssymb,amsfonts}       % Typical maths resource packages

\usepackage{hyperref}%

\usepackage{color}
\newcommand{\average}[1]{\left \langle {#1} \right \rangle}

\newcommand{\T}{\mathsf{T}} % matrix transpose
\newcommand{\0}{\mathsf{0}} % null matrix 
\newcommand{\I}{\mathbf{I}} % identity matrix 
\newcommand{\Proj}{\mathbf{H}} % Projector
\newcommand{\CovMat}{\boldsymbol{\Sigma}} % reconstructor

\newcommand{\N}{\mathbf{N}} % DM influence function operator
\newcommand{\D}{\mathbf{G}} % WFS  operator
\newcommand{\dint}{\mathrm{d}} % var intégration 
 
\newcommand{\cor}{\mathsf{\tiny{cor}}}

\newcommand{\thetavec}{{\boldsymbol{\theta}}} 
\newcommand{\alphavec}{{\boldsymbol{\alpha}}} 
\newcommand{\betavec}{{\boldsymbol{\beta}}} 
\newcommand{\rhovec}{\boldsymbol{\rho}}%{{\boldsymbol{\rho}}} 
\newcommand{\etavec}{{\boldsymbol{\eta}}} 
\newcommand{\phivec}{{\boldsymbol{\psi}}} 
\newcommand{\varphivec}{{\boldsymbol{\varphi}}} 
 
\newcommand{\svec}{{\mathbf{s}}} 
\newcommand{\xvec}{{\mathbf{x}}} 
\newcommand{\vvec}{{\mathbf{v}}} 
\newcommand{\uvec}{{\mathbf{u}}} 
 
\newcommand{\Asa}{{\mathbf{A}}}
\newcommand{\Aexp}{{\mathbf{A}_\text{\tiny{E}}}}
\newcommand{\AexpT}{{\mathbf{A}_\text{\tiny{E}}^\T}}

\newcommand{\Mexp}{{\mathcal{M}_\infty^\text{\tiny{E}}}}
\newcommand{\Lexp}{{\mathcal{L}_\infty^\text{\tiny{E}}}}
\newcommand{\Sigmaexp}{{\CovMat_\infty^\text{\tiny{E}}}}
\newcommand{\SigmaexpN}{{\CovMat_n^\text{\tiny{E}}}}
\newcommand{\Msa}{{\mathcal{M}_\infty}}
\newcommand{\SigmaSA}{{\CovMat_\infty}}
\newcommand{\SigmaSAN}{{\CovMat_n}}

\newcommand{\Ndm}{{{N_\beta}}} 
\newcommand{\Nact}{{{N_\text{act}}}} 
\newcommand{\Nwfs}{{{N_\alpha}}} 
\newcommand{\Nslopes}{{{N_\text{s}}}} 
\newcommand{\Nl}{{{N_\text{l}}}} 
\newcommand{\Nphi}{{{N}_\varphi}} 

\newcommand{\qed}{\nobreak \ifvmode \relax \else
      \ifdim\lastskip<1.5em \hskip-\lastskip
      \hskip1.5em plus0em minus0.5em \fi \nobreak
      \vrule height0.75em width0.5em depth0.25em\fi}

\begin{document}

%%%%%%%%%%%%%%%%%% title page information %%%%%%%%%%%%%%%%%%
\title{Spatio-angular Minimum-variance Tomographic Controller for Multi-Object Adaptive Optics systems} 
%\author{C. Correia and K. Jackson}
\author{Carlos M. Correia}\email{carlos.correia@lam.fr} %% email address is required
\affiliation{Aix Marseille Universit\'e, CNRS, LAM (Laboratoire d'Astrophysique de Marseille) UMR 7326, 13388, Marseille, France}
%\affiliation{Dept. of Physics and Astronomy, Faculty of Sciences, University of Porto, Rua do Campo Alegre 687, PT4169-007 Porto, Portugal}
%\affiliation{Centre for Astrophysics, University of Porto, Rua das Estrelas, 4150-762 Porto, Portugal}
\affiliation{Adaptive Optics Laboratory, University of Victoria,
  Victoria, BC V8P 5C2}
\affiliation{Formerly with the Institute of Astrophysics and Space Sciences, University of Porto, CAUP, Rua das Estrelas, 4150-762 Porto, Portugal}
\author{Kate Jackson}
\affiliation{Adaptive Optics Laboratory, University of Victoria,
  Victoria, BC V8P 5C2}
\affiliation{Dept. of Mechanical Engineering, University of Victoria, Canada}
\author{Jean-Pierre V\'eran}
\affiliation{National Research Council, Herzberg Institute of
  Astrophysics, 5071 West Saanich Road, Victoria, British Columbia V9E
  2E7, Canada}
\author{David Andersen}
\affiliation{Adaptive Optics Laboratory, University of Victoria,
  Victoria, BC V8P 5C2}
\affiliation{National Research Council, Herzberg Institute of
  Astrophysics, 5071 West Saanich Road, Victoria, British Columbia V9E
  2E7, Canada}
\author{Olivier Lardi\`ere}
\affiliation{Adaptive Optics Laboratory, University of Victoria,
  Victoria, BC V8P 5C2}
\author{Colin Bradley}
\affiliation{Adaptive Optics Laboratory, University of Victoria,
  Victoria, BC V8P 5C2}
\affiliation{Dept. of Mechanical Engineering, University of Victoria, Canada}
%\author{(C$^2$)\,$^{a,b,*}$, Kate Jackson$^b$, Jean-Pierre V\'eran$^a$ \textit{et al}}
% \affiliation{$^{a}$ Centre for Astrophysics, University of Porto, Rua das Estrelas, 4150-762 Porto, Portugal}
% \affiliation{$^{b}$Adaptive Optics Laboratory, University of Victoria, Victoria, BC V8P 5C2}
% \address{$^{c}$Herzberg Institute of Astrophysics, National Research Council, Canada}
%\\$^{*}$carlos.correia@nrc.gc.ca}
 
%\address{ $^{b}$AO lab, UVic, Victoria, BC Canada} 
%\email{$^*$ccorreia@astro.up.pt} %% email address is required

%\homepage{http://astro.up.pt/~ccorreia/} %% author's URL, if desired


\begin{abstract}
%\textit{Context:} 
Multi-object astronomical adaptive-optics (MOAO) is now a mature wide-field observation mode to enlarge the adaptive-optics-corrected field in a few specific locations over tens of arc-minutes. 

%\textit{Aims:} 
The work-scope provided by open-loop tomography and
pupil conjugation is amenable to a spatio-angular Linear-Quadratic
Gaussian (SA-LQG) formulation aiming to provide enhanced correction across the
field with improved performance over static reconstruction methods and
less stringent computational complexity scaling laws. 

%\textit{Methods:} 
Starting from  our
previous work \cite{correia14}, we use stochastic time-progression models coupled to
approximate sparse measurement operators to outline a suitable SA-LQG formulation capable of delivering near optimal correction. Under the spatio-angular framework the wave-fronts are never explicitly estimated in the volume, providing considerable computational savings on 10\,m-class telescopes and beyond. 

%\textit{Results:} 
We find that for Raven, a 10m-class MOAO system with two science channels, the SA-LQG improves the limiting magnitude by two stellar magnitudes when both Strehl-ratio and Ensquared-energy are used as figures of merit. 
The sky-coverage is therefore improved by a factor of $\sim$5. 

% \ccc{The computational savings are such that on a ELT-sized system with 20 parallel targets we foresee up to 5 times the number of operations only!}
% Abstract...
% % Generally called wide-field adaptive optics systems require
% %   tomographic reconstruction of wave-front phase in the
% %   atmosphere. In recent years, different flavours of (static) tomographic reconstructors
% %   were made available that suit more specifically a particular AO
% %   setup. 
\end{abstract}
 % REPLACE WITH CORRECT OCIS CODES FOR YOUR ARTICLE

%%%%%%%%%%%%%%%%%%%%%%% References %%%%%%%%%%%%%%%%%%%%%%%%%
\maketitle 

%\bibliographystyle{osajnl}   %>>>> makes bibtex use osajnl.bst
%\bibliography{references}   %>>>> bibliography data in references.bib

% \begin{thebibliography}{99}

% \bibitem{gallo99} K. Gallo and G. Assanto, ``All-optical diode based on second-harmonic generation in an asymmetric waveguide,'' \josab {\bf 16,} 267--269 (1999).

% \end{thebibliography}

% \textbf{TODO}

% 1. Plot the mode-by-mode variance for direction thetaii using the
% spatio-angular and the projection of the explicit reconstructor as a
% function of the projected radial orders.

% 2. Compare the SVD of the full spatio-angular Caa with the Caa
% obtained by projections

% 2.5 Repeat process for zonal representation

% 3. Compare the above matrices with the ones collected from telemetry
% (as a function of the number of averaged time-steps)

% 4. Discuss the advantages of having prediction in the layers. Assess
% finally the effectiveness of the AR1/AR2 models for open-loop
% phase-prediction compared with the frozen-flow (say estimate the
% 1-step estimation error from using AR instead of screen shifting)

%\newpage
%%%%%%%%%%%%%%%%%%%%%%%%%%%%%%%%%%%%%%%% 
% ------------section--------------------
%%%%%%%%%%%%%%%%%%%%%%%%%%%%%%%%%%%%%%%% 
\section{Intro: Multi-object adaptive optics systems}

Multi-object Adaptive Optics (MOAO) is now a well-established AO concept to enlarge the corrected field to several arcminutes when only a few directions of interest exist in the field and need be corrected for \cite{vidal10, andersen12, correia14, sivo14}. 

Similarly to multi-conjugate AO (MCAO) it relies on tomography to provide a tridimensional wave-front estimation above the telescope and thereupon provide correction on several scientific relevant directions, reaching out for a multiplexing factor of up to 20 on ELT-sized systems \cite{andersen06, hammer14}.

On a previous paper \cite{correia14} we have outlined the static
\textit{minimum-mean square error} (MMSE) \textit{spatio-angular} (SA)
wave-front estimation (time-dependent wave-front evolution not
considered) when a Zernike polynomial basis is used and suggested predictive models to overcome temporal lag errors when aiming for increased sky-coverage \cite{jackson15}. In the SA case estimating the tomographic phase explicitly is circumvented, only the aperture-plane is estimated to be later least-squares fitted onto the DM -- the \textit{Learn\&Apply} approach in \cite{vidal10} is similar albeit in a different vector space. 

We now move one step forward: we provide extensions to use a zonal
basis set coupled to the optimal minimum residual phase error variance
Linear-Quadratic-Gaussian (LQG) controller tailored to MOAO
systems.The LQG controller has been presented in the AO context in
several works \cite{piatrou05, correia10a, gilles13, sivo14} and here it is designed and optimized for a MOAO system performing open-loop tomography and single-layer optical conjugation. We further propose a minimal state representation together with first-order near-Markovian predictive models which lead to a computational complexity of the order of the static MMSE reconstructors for a multiplexing capability of 20 science targets. 

The use of a spatio-angular framework has several advantages over the formulation that estimates the layered phase explicitly: 
\begin{enumerate}
\item the real-time computation is independent from the number of atmospheric layers; 
\item  the analytical spatio-angular covariance matrices can be computed explicitly thus avoiding interpolation through the ray-tracing operation to obtain the integrated pupil-plane wave-front profile;
\item in principle a hybrid data-driven model-based controller can be designed since a subset of the  covariance matrices can be acquired directly from the system (i.e. empirically acquired) instead of relying fully on models; alternatively meta-parameters can be identified from which enhanced models are obtained.
\item the sparseness of the LQG compounding matrices can still, albeit
  to a lesser degree, be exploited for real-time savings as noted in
  \cite{jackson15, correia09}.
\end{enumerate}
% However, the sparseness of the resulting matrices cannot be exploited for simulation and real-time savings as advocated for instance in \cite{ellerbroek02, gilles13, rosensteiner11} and others. We note however that advances in memory handling and chipset clock speeds may move away from sparse implementations in real-time controllers \cite{veran14}.

The goal of this paper is therefore two-fold: 

\textit{i}) fully outline the zonal LQG minium-variance tomographic
controller design to optimize optical performance, particularly in
poor signal-to-noise ratio regimes such that sky-coverage can be improved upon by guiding on fainter sources %in line with previous work in \textit{Correia et al} \cite{correia14} 
and 

\textit{ii}) provide a suitable, \textit{Extremely Large Telescope} (ELT)-compliant spatio-angular formulation for the Kalman estimator (and static MMSE) that decouples the wave-front reconstruction from the number of atmospheric layers used to model it. 

We present results from both simulation and the optical bench of the Raven system, a MOAO science and technology demonstrator installed on the 8\,m Subaru telescope \cite{andersen12}.
% We provide numerical assessment for the  Raven system. Raven is a MOAO science and technology demonstrator installed on the 8\,m Subaru telescope. Moreover, we provide empirical results from the bench while we gear up for on-sky tests.

This paper is organized as follows. Section 2 outlines the LQG formulation, section 3 discusses other SA implementations, namely the static SA reconstructors obtained from the LQG, section 4 provides sample numerical results whereas section 5 overviews the Raven optical bench and the results obtained in the lab. Section 6 provides the summary and conclusions.  

%%%%%%%%%%%%%%%%%%%%%%%%%%%%%%%%%%%%%%%% 
% ------------section--------------------
%%%%%%%%%%%%%%%%%%%%%%%%%%%%%%%%%%%%%%%% 
\section{Spatio-angular LQG formulation}

\subsection{Definitions and assumptions}\label{sec:definitions}
We define the pupil-plane wave-front  $\phivec(\rhovec,\thetavec,t) $ indexed by the bi-dimensional spatial coordinate vector $\rhovec = [\rho_x, \rho_y]$ in direction $\thetavec = [\theta_x, \theta_y]$ at time $t$ under the hypothesis that the turbulent atmosphere is a sum of $L$ thin layers located in a discrete number of different altitudes $h_l$ as
 \begin{equation}
\phivec(\rhovec,\thetavec,t) = [\Proj \varphivec] (\rhovec,t) = \sum_{l=1}^{L} \omega_l \varphivec_l(\rhovec + h_l \thetavec,t) 
\end{equation}
where $\varphivec_l(\rhovec,t)$ is the $l^{th}$-layer wave-front,
$\omega_l$ is the $l^{th}$ layer relative strength and $\Proj$ is a
geometric optics propagation operator in the near-field approximation
(commonly called ray-tracing  \cite{gilles02}) that relates the
aperture-plane wave-front (WF) to the layered wave-fronts by adding and
interpolating $\varphivec_l(\rhovec + h_l \thetavec,t)$ on the
aperture-plane ($h_0 = 0$) computational grid \cite{ellerbroek02a}. 

In the following we assume the frozen-flow hypothesis whereby $\varphivec(\boldsymbol{\rho},t+\tau) = \varphivec(\boldsymbol{\rho} - \mathbf{v}\tau,t)$ with $\mathbf{v} = [v_x; v_y]$ the wind velocity vector. A discrete-time, point-wise representation of the WF is also assumed throughout this paper.

% We start our presentation by adopting the discrete-time representation
% and 
We further assume that averaged phase values over the wave-front
sensor (WFS) integration period $T_s$ are taken at face value, \textit{i.e} 
\begin{equation}
\varphivec_k(\rhovec) = \frac{1}{T_s}\int_{(k-1)T_s}^{k T_s}\varphivec(\rhovec, t)\, \dint t,
\end{equation}
and that the commands are piece-wise constant over an entire frame (either synchronous or not with the WFS sampling frames)
\begin{equation}
\uvec_k \triangleq \uvec(t), \,\,\,\, kT_s \leq t < (k+1) T_s 
\end{equation}

Ensemble averaging is represented by $\average{\cdot}$ with shorthand covariance matrix notation $\boldsymbol{\Sigma}_{u,v} = \average{u v^\T}$. Angles are notated in general $\thetavec$ and guide-star and science directions use $\alphavec$ and $\betavec$ respectively.

\subsection{LQG regulator synthesis}
%\subsection{Kalman filter models and design}
The derivation of the LQG for AO applications has been presented in detail elsewhere for either the infinitely fast DM response or otherwise \cite{correia10a}. In this paper we therefore take a shorter path for the sake of conciseness and refer the reader to the bibliographic references on this particular subject for a more in-depth derivation. 

The discrete-time LQG regulator minimizes the cost function
\begin{align}
J(\uvec) & = \lim_{M \rightarrow \infty}\frac{1}{M}\sum_{k=0}^{M-1} \left(\xvec^\T \mathbf{Q} \xvec  + \uvec^\T \mathbf{R} \uvec + 2 \xvec^\T \mathbf{S} \uvec \right)_k
\end{align}
where the weighting triplet $\{\mathbf{Q}, \mathbf{R}, \mathbf{S}\}$ is made apparent and will be specified next by developing a MOAO-specific quadratic criterion % in Eq. (\ref{eq:Ju}) along with the state-space matrices $\{ \mathcal{A}, \mathcal{B}, \mathcal{V}, \mathcal{C}, \mathcal{D}\}$ after the definition of phase time-evolution and measurement equations.
subject to the state-space model
\begin{equation}\label{eq:state_space_model}
  \left\{\begin{array}{ccl}
     \xvec_{k+1}  &  = & \mathcal{A} \xvec_k  + \mathcal{B} \uvec_k  + \mathcal{V}  \boldsymbol{\nu}_k \\
     \svec_{\alphavec,k}& = & \mathcal{C} \xvec_k + \mathcal{D} \uvec_k  + \etavec_k
    \end{array}\right.
\end{equation}
where $\xvec_k$ is the state vector that contains if not the phase directly a linear combination thereof, $\svec_{\alphavec,k}$ are the noisy measurements provided by the WFS in the GS directions, $\boldsymbol{\nu}_k$ and $\etavec_k$ are spectrally white, Gaussian-distributed state excitation and measurement noise respectively. In the following we assume that $\CovMat_{\boldsymbol{\nu}} = \average{\boldsymbol{\nu} \boldsymbol{\nu}^\T}$,  $\CovMat_{\mathbf{\boldsymbol{\eta}}} = \average{\boldsymbol{\eta} \boldsymbol{\eta}^\T}$ are known and that $\CovMat_{\mathbf{\boldsymbol{\eta,\nu}}} = \average{\boldsymbol{\eta} \boldsymbol{\nu}^\T} = 0$.

% The controller synthesis relies on the separation principle between control and estimation \cite{andersonmoore_optimalcontrolLQG05}. Consequently, we  compute the negative-feedback regulator assuming a know state and estimate the state from a Kalman filter. 

\subsection{Minimum residual phase variance after correction}
The objective cost function considered for MOAO is the minimisation of the
aperture-plane residual
phase variance for individual science directions $\betavec_i \in \Re^{1\times 2} $ leading to the maximisation of the Strehl-ratio. 
% , $\varphivec$ represents the wave-front profiles in the atmospheric layers at any altitude whereas $\Proj$ stands for a ray-tracing and bilinear interpolation operation along the science objects lines of sight; $W$ is a positive-definite weighting matrix that removes the piston contribution over the telescope aperture.

If we denote the term $\boldsymbol{\epsilon}_k \triangleq {\phivec}(\rhovec, {\beta_i},kT_s) - {\phivec}^\cor(\rhovec, {\beta_i},kT_s)$ the residual phase profile after DM correction with ${\phivec}(\rhovec, {\beta_i},kT_s) =\Proj_{\beta_i} {\varphivec}(\rhovec, kT_s)$ the aperture-plane wave-front in a particular direction in the field and $- {\phivec}^\cor(\rhovec, {\beta_i},kT_s)$ the DM-produced correction phase in that direction, the mean square residual piston-removed wave-front error over the telescope aperture 
 is $\sigma^2_k = \Vert \boldsymbol{\epsilon}_k \Vert^2_{W,L_2(\Omega)} = \boldsymbol{\epsilon}_k^\T W \boldsymbol{\epsilon}_k$ where  $W$ is a positive-definite weighting matrix that removes the piston contribution over the telescope aperture \cite{ellerbroek02} and $L_2$ is the Euclidean norm over the aperture $\Omega$.

% The discrete-time nature of the AO loop is prone to delays leading to temporal lag errors intrinsic to the system like detector integration, read-out and reconstruction which take customarily several milliseconds.

The integration, reading, processing and correction in the AO loop are done in discrete-time leading to intrinsic lags which take customarily several milliseconds. We will expand the discussion of time delays for synchronous and asynchronous systems in \S \ref{sec:time_delayed_correction}.

Using the definitions of integrated phase and commands in \S \ref{sec:definitions} we can now ascertain the discrete-time minimum residual phase variance (MV) criterion
\begin{align}\label{eq:Ju}
J(\uvec) & = \lim_{M \rightarrow \infty}\frac{1}{M}\sum_{k=0}^{M-1} \sigma^2_k \nonumber \\
& =  \lim_{M \rightarrow \infty}\frac{1}{M}\sum_{k=0}^{M-1}\Vert
  \phivec_{k+1} (\rhovec,\betavec) - \N \uvec_k  (\rhovec,\betavec) \Vert^2_{W,L_2(\Omega)}
\end{align}
which neglects any WF dynamics during the integration. The error that is neglected here has been dubbed \textit{insurmountable error} due to the use of averaged variables instead of continuous ones \cite{correia10a} and is negligible for the most common systems operating at high frame-rates.

Provided the phase vector ${\phivec}_{k+1}$ is known the solution to Eq. (\ref{eq:Ju}) is straightforwardly shown to be a least-squares fit onto the DM influence functions \cite{kulcsar06, correia10a}
\begin{align}\label{eq:CSI}
\uvec_k  (\rhovec,\betavec) & = \mathbf{F}{\phivec}_{k+1}
                              (\rhovec,\betavec) \nonumber\\& = (\N^\T W\N)^{-1}\N^\T W \phivec_{k+1} (\rhovec,\betavec)%\widehat{\phivec}_{k+1|k}
%\uvec = \mathbf{F} \widehat{\phivec}_{k+1|k}
\end{align}
The latter would have been likewise found from the degenerate control Riccati equation, whose discussion we avert here but can be found for instance in \cite{kulcsar12, correia10a}.

We now deal with the general case where $\phivec_{k+1}$ is not known in Eq. (\ref{eq:CSI}) but needs be estimated  from linearly related noisy measurements.

The separation principle \cite{andersonmoore_optimalcontrolLQG05} provides immediately the solution  
\begin{align}
\uvec_k  (\rhovec,\betavec)  & = \mathbf{F}\widehat{\phivec}_{k+1|k} (\rhovec,\betavec) \nonumber \\
& = (\N^\T W\N)^{-1}\N^\T W \widehat{\phivec}_{k+1|k} (\rhovec,\betavec),
%\uvec = \mathbf{F} \widehat{\phivec}_{k+1|k}
\end{align}
where $\widehat{\phivec}_{k+1|k} (\rhovec,\betavec) =
\widehat{\phivec}_{k+1|\mathcal{S}_k(\alphavec)}  (\rhovec,\betavec)=
E({\phivec}_{k+1} (\rhovec,\betavec)|\mathcal{S}_k(\rhovec,\alphavec))$ is the conditional expectation of
${\phivec}_{k+1} (\betavec)$ in the science directions  with respect to the sequence of all
measurements from (in general non-coincidental) GS directions available up to $t=k T_s$, \textit{i.e.} $\mathcal{S}_k =
\{\svec_0, \cdots,\svec_k\}(\rhovec,\alphavec)$ 
\cite{andersonmoore_optimalfiltering05}. This conditional expectation
is readily available from a Kalman filter
\cite{andersonmoore_optimalcontrolLQG05, correia10a, gilles13}.  It
relies on the definition of the first-differences state-space model
in Eq. \eqref{eq:state_space_model} whose construction is shown
further below. %  which involves defining the following state-space model 
% \begin{equation}
%   \left\{\begin{array}{ccl}
%      \xvec_{k+1}  &  = & \mathcal{A} \xvec_k  + \mathcal{B} \uvec_k  + \mathcal{V}  \boldsymbol{\nu}_k \\
%      \svec_{\alphavec,k}& = & \mathcal{C} \xvec_k + \mathcal{D} \uvec_k  + \etavec_k
%     \end{array}\right.
% \end{equation}
% where $\xvec_k$ is the state vector that contains if not the phase directly a linear combination thereof, $\svec_{\alphavec,k}$ are the noisy measurements provided by the WFS in the GS directions, $\boldsymbol{\nu}_k$ and $\etavec_k$ are spectrally white, Gaussian-distributed state excitation and measurement noise respectively. 

% Synthesis of the Kalman filter relies on the minimisation of a quadratic criterion of the form
% \begin{align}
% J(\uvec) & = \lim_{M \rightarrow \infty}\frac{1}{M}\sum_{k=0}^{M-1} \left(\xvec^\T \mathbf{Q} \xvec  + \uvec^\T \mathbf{R} \uvec + 2 \xvec^\T \mathbf{S} \uvec \right)_k
% \end{align}
% where the weighting triplet $\{\mathbf{Q}, \mathbf{R}, \mathbf{S}\}$ is made apparent and will be specified next by developing the quadratic criterion in Eq. (\ref{eq:Ju}) along with the state-space matrices $\{ \mathcal{A}, \mathcal{B}, \mathcal{V}, \mathcal{C}, \mathcal{D}\}$ after the definition of phase time-evolution and measurement equations.

\subsubsection{near-Markovian time-progression model of the 1st-order }\label{sec:nearMarkovModel}
The use of auto-regressive models has been given wide attention using
the Zernike polynomials' expansion basis set \cite{petit09, sivo14} as
well as with Fourier modes for which a complex model encodes perfectly
spatial shifts and thus frozen-flow~\cite{poyneer08}. %Several time-progression models have been analysed over the past few years. % Predictive schemes exploring the frozen-flow are most common either embedding layer disentanglement \cite{poyneer08} or not \cite{lukin10}.

We restrict our attention to the near-Markovian first-order
time-evolution model \cite{gavel02a} which has delivered best overall
performance albeit with increased computational
complexity~\cite{piatrou07, correia14, jackson15} (spatial dependence omitted)
\begin{equation}\label{eq:AR1model}
  \phivec_{k+\Delta}(\alphavec) = \Asa \phivec_{k} (\alphavec) + \boldsymbol{\nu}_k (\alphavec)
\end{equation}  
with  $\Delta = \tau/T_s$ is a delay in units of sample step $T_s$, where the transition matrix minimizes the quadratic criterion \cite{correia14}
\begin{equation}\label{eq:Asa}
 \Asa = \arg\min_{\Asa'}\average{\left\Vert\phivec_{k+\Delta} (\alphavec)- \Asa'  \phivec_k(\alphavec) \right \Vert^2_{L_2(\Omega)}},
\end{equation}

The solution to Eq. (\ref{eq:Asa}) is 
\begin{equation}\label{Asa_Def}
\Asa = \average{\phivec_{k+\Delta}\phivec_k^\T } \average{ \phivec_k \phivec_k^\T }^{-1} (\alphavec)
\end{equation}
Note that we do not remove piston in this equation although we could following formulae in \cite{wallner83}.

\begin{figure}[htpb]
	\begin{center}
	% use packages: array            
        %  \begin{tabular}{cc}
            \includegraphics[width=1.0\textwidth]{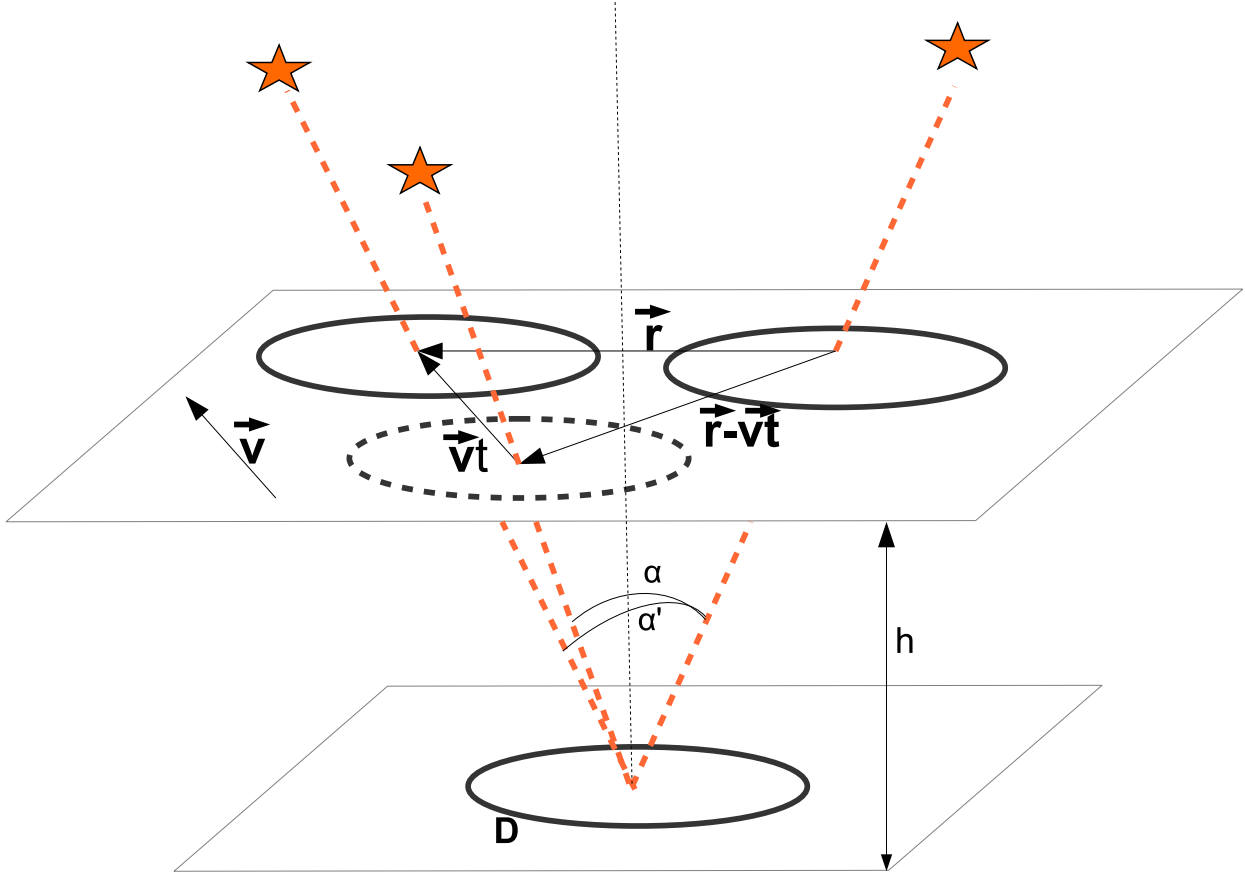}
         % \end{tabular}
	\end{center}
	\caption[]
	{\label{fig:on_sky_angles}
Spatio-angular computation scheme. Pupil displacements are highly exaggerated for viewing. The spatio-angular correlations are computed with the pupil located at $\rhovec-\vvec \tau$ where $\tau$ is time in seconds.}
\end{figure}

% from which 
% \begin{equation}\label{eq:zonal_A_matrix}
%   \mathcal{A} = \average{\phivec_{k+\Delta}\phivec_k^\T }
%   \average{ \phivec_k \phivec_k^\T }^{-1} = \Proj \average{\varphivec_{k+\Delta}\varphivec_k^\T }
%   \average{ \varphivec_k \varphivec_k^\T }^{-1} \Proj^{-1},
% \end{equation}
By denoting $\phivec_{k}(\rhovec, \alphavec) = \Proj_\alphavec\varphivec(\rhovec, k T_s)$ and substituting in Eq. (\ref{Asa_Def}) we get 
% Under the Taylor frozen-flow hypothesis, Eq.~(\ref{eq:Asa}) is specified as
\begin{align}\label{eq:Atur}
  \Asa &  = \Proj_\alphavec \average{\varphivec_{k+\Delta}\varphivec_k^\T }
  \average{ \varphivec_k \varphivec_k^\T }^{-1} \Proj_\alphavec^{\dag}  \\
&  = \Proj_\alphavec \average{\varphivec(\rhovec - \mathbf{v}\tau)\varphivec(\rhovec)^\T }
  \average{ \varphivec(\rhovec) \varphivec(\rhovec)^\T }^{-1} \Proj_\alphavec^{\dag} \nonumber 
%%\\& = \average{\phivec_{k}(\rhovec- \mathbf{v}\tau)\phivec_{k}(\rhovec)^\T }  \average{ \phivec_{k}(\rhovec)^\T\phivec_{k}(\rhovec)^\T }^{-1} 
\end{align}
where  $\phivec_{k+\Delta}(\rhovec, \alphavec) = \Proj_\alphavec \varphivec(\rhovec- \mathbf{v}\tau kT_s) = \Proj_{\alphavec'}\varphivec(\rhovec, kT_s)$  % and $\phivec_k = \phivec(\rhovec, \alphavec) = \phivec(\alphavec, kT_s)$ ,
which translates the spatio-angular nature of the computation we are accomplishing since a time-translation is factored in as an altitude-dependent angular shift using the frozen-flow hypothesis. Figure \ref{fig:on_sky_angles} shows pictorially the statements made.

Assuming stationarity, the state excitation noise covariance matrix is found for a first order time-evolution model
from the covariance equality (implicit indices dropped out)
 \begin{equation}\label{eq:state_noise_matrix}
   \average{\phivec \phivec^\T } = \Asa
   \average{ \phivec \phivec^\T }
   \Asa^\T + \average{\boldsymbol{\nu} \boldsymbol{\nu}^\T},
 \end{equation}
since $\average{\phivec_{k+1}\phivec_{k+1}^\T } = \average{ \phivec_{k} \phivec_{k}^\T } = \average{\phivec_{}\phivec_{}^\T }$. The excitation noise covariance matrix is therefore
\begin{equation}
\average{\boldsymbol{\nu} \boldsymbol{\nu}^\T} = \average{ \phivec \phivec^\T } - \Asa \average{ \phivec \phivec^\T } \Asa^\T
 \end{equation}  
The model driving noise covariance matrix $\average{\boldsymbol{\nu} \boldsymbol{\nu}^\T}$ is a key element of the KF design.
% Note however that Eq. (\ref{eq:Atur}) is never explicitly computed. Section \ref{sec:CompCovMat} provides the method to compute $\Asa$ directly as
% \begin{equation}
% \Asa = \average{\phivec_{k+\Delta}\phivec_k^\T } \average{ \phivec_k \phivec_k^\T }^{-1} 
% \end{equation}

% and replacing $\mathcal{A}^{*}_\alphavec$ by its formulation one ends
% up with 
%          \begin{equation}\label{eq:zonal_A_matrix}
%            \Sigma_v =   \average{ \phivec_{k,\alphavec}
%              \phivec_{k,\alphavec}^\T } -
%            \average{\phivec_{k+1,\alphavec}\phivec_{k+1,\alphavec}^\T }  \average{ \phivec_{k,\alphavec}
%              \phivec_{k,\alphavec}^\T }^{-1}\average{\phivec_{k+1,\alphavec}\phivec_{k+1,\alphavec}^\T } 
%          \end{equation}

%%%%%%%%%%%%%%%%%%%%%%%%%%%%%%%%%%%%%%%% 
% ------------section--------------------

\subsubsection{Measurement model}
The presentation that now takes place restricts itself to the case of the Shack-Hartmann (SH) WFS which provides the averaged phase gradient over a regular grid of sub-apertures in the pupil-plane \cite{hardy98}.

We use a geometrical linear model which we assume valid for the regime of operation it will be submitted to; the model is thus
\begin{equation}\label{meas_model}
\svec_{k}(\alphavec) = \D  \phivec_{k}(\alphavec) + \etavec_k(\alphavec)
\end{equation}
where $\etavec_k$ represents photon an detector shot noise taken to be additive Gaussian noise of know covariance and $\D$ is a matrix whose entries are computed from the discretization of the wave-front gradients in the $\Nwfs$ GS directions. 
It uses a 3x3 regular stencil to compute the averaged phase gradient with Simpson weights for the fully-illuminated sub-apertures
\begin{equation}
w_x = \frac{1}{d}
\left[
\begin{array}{ccc}
-1/4 & 0 & 1/4 \\
-1/2 & 0 & 1/2 \\
-1/4 & 0 & 1/4
\end{array}\right], w_y = -w_x^\T
\end{equation}
The stencils for partially illuminated sub-apertures at the outer and inner edge of the telescope pupil are found by expanding the phase on a set of bilinear splines and computing the average gradient over each of the four quadrant patches. The derivation of this result is out of scope but can be found at request in \cite{waddle12}.

This sampling and respective estimation of wave-fronts at twice the SH-WFS's Nyquist frequency permits the rejection of some aliasing that affects the original measurements as outlined in \cite{petit09}. An enhanced model with coloured noise is shown in \cite{poyneer10}.  

% Given that all points in the aperture grid do not receive the same amount of illumination,
% the computation of an average gradient in a subaperture must include a weighting by the
% squared amplitudes of the illumination received by each of its nine grid points. The stencils for the partially illuminated sub-apertures are found by expanding the phase on a set of bilinear splines and computing the average gradient over each of the four quadrant patches. The derivation of this result is out of scope but can be found at request in \cite{waddle12}.

% Matrix $\Proj_\alphavec$ represents ray-tracing whereby rays falling on each and every point in the aperture-grid along direction $\alphavec$ are intersected with phase screens in the atmosphere and points linearly interpolated and accumulated \cite{gilles02}. 

Note that in our implementation we will not make explicit use of the ray tracing $\Proj$ in Eq. (\ref{meas_model}) since the layered phase need not be estimated -- no multi-conjugation step ensues. Only the pupil-plane phase is of interest which opens up to considerable modelling simplifications using a spatio-angular frame-work. 
% Therefore the measurement equation collapses to
% \begin{equation}
% \svec_{\alphavec,k} = \D  \phivec_{\alphavec,k} + \etavec_k
% \end{equation}
% where $\phivec_{\alphavec,k} = \Proj_\alphavec\varphivec_k $.

This is a fundamental simplification with respect to previous
tomographic LQG implementations -- which in the remainder is called
E-LQG to account for the explicit estimation step. We avoid estimating
the layered wave-front as is customarily done, focusing in estimating the integrated pupil-plane wave-fronts in the GS directions along their lines of sight. The angular anisoplanatic estimate is made on a subsequent step by invoking linearity, stationarity and Gaussianity as in the static case \cite{correia14}. % Opting for a reduced size vector space leads to considerable computational savings as will be shown next.

\subsection{Minimal state-space representation: compactness in Open-Loop AO}

The combination of enhanced spatio-angular first-order predictive models with single-conjugate correction allows for some modelling simplifications that in turn will make for smaller problem size and less stringent real-time and off-line computational requirements. 

Under these working assumptions, the LQG model admits -- unlike previous implementations \cite{piatrou07,sivo14} -- a minimal state representation with phase taken at one single time instant.  To do so one assumes the measurements are available at the end of the WFS integration step, regardless of the actual temporal delay in the system \cite{correia10a}: one affects thus the whole delay to the commands instead of the measurements.

On top of this, we can further use a state vector with
% We now make a further fundamental simplification: instead of defining the state vector as phase in the layers as is customarily done, we use
the pupil-plane phase integrated over directions $\alphavec \in[1, \cdots, N_\alpha]$ at instant $k$ to provide the \textit{Spatio-Angular} LQG formulation. 

Selecting thus $\xvec_{k}\triangleq \phivec_{k}(\alphavec)$ 
\begin{equation}
\phivec_{k}(\alphavec) =
\left(\begin{array}{c}
 \phivec({\alphavec_1})\\ \cdots\\ \phivec({\alphavec_{N_\alpha}})
\end{array}\right)_k
\end{equation}
where the state is a concatenation of phase vector in the $N_\alpha$ guide-star directions, 
one defines the state space terms for Eq. (\ref{eq:AR1model})
\begin{equation}\label{eq:stack_ss_c_tur_compact}
  \left[\begin{array}{c|c|c}
      \mathcal{A} & \mathcal{B} &  \mathcal{V}\\\hline
      \mathcal{C} & \mathcal{D}  & \0
    \end{array}\right]
=
\left[\begin{array}{c|c|c}
      \Asa & \0 & \I \\\hline
      \D & \0 & \0
    \end{array}\right]
\end{equation}

The implementation of the KF involves a real-time state update and prediction equations which in the SA case are
\begin{subequations}\label{eq:single_rate_StatSA_RToperations}
\begin{align}
  \widehat{\phivec}_{k|k}(\alphavec) & = \widehat{\phivec}_{k|k-1}(\alphavec) + \mathcal{H}_\infty\left(\svec_k(\alphavec) -
     \D \widehat{\phivec}_{k|k-1}(\alphavec) \right) 
\\
  \widehat{\phivec}_{k+1|k}(\alphavec) & = \Asa\widehat{\phivec}_{k|k} (\alphavec)
\end{align}
\end{subequations}
where $\mathcal{H}_\infty$ is the asymptotic Kalman gain computed from the solution of an estimation Riccati equation \cite{correia10a}. The use of the asymptotic value is justified, like elsewhere \cite{petit08} for one seeks long-exposure performance therefore employing the steady-state gain with no loss of performance. Whenever implicit we drop the notation $(\alphavec)$.

%The mathematical relationship of the SA-LQG and the explicit LQG is provided in Appendix \ref{app:SA_vs_Explicit_LQG}.

Under the choice of state made earlier, the $L_2$ criterion in
Eq. (\ref{eq:Ju}) assumes the equivalent form 
\begin{align}
J(\uvec) = & \lim_{M \rightarrow \infty} \nonumber \\ & \frac{1}{M}\sum_{k=0}^{M-1} 
\left(
\begin{array}{c}
  \xvec_{k}\\\uvec_k
\end{array}\right)^\T
\left(
\begin{array}{cc}
   \Asa^\T\Asa & -\Asa^\T\N \\ -\N^\T\Asa & \N^\T \N
\end{array}\right)
\left(
\begin{array}{c}
  \xvec_{k}\\\uvec_k
\end{array}\right)
%\xvec^\T \mathbf{Q} \xvec  + \uvec^\T \mathbf{R} \uvec + 2 \xvec^\T \mathbf{S} \uvec \right)
\end{align}
from which $\mathbf{Q} = \Asa^\T \Asa\geq \mathbf{0}$, $\mathbf{R} = \N^\T\N > \mathbf{0}$ and $\mathbf{S} = -\Asa^\T\N$; this construction ensures a stable controller since $\Asa$ is a positive definite stability matrix with $|eig( \Asa)|<1$ and $\mathbf{Q} - \mathbf{S}\mathbf{R}\mathbf{S}^\T\geq 0$ and $\mathbf{R}$ a full-rank matrix.
%\ccc{Could you check the inequalities above please? }

\subsection{Time-delayed correction: synchronous and asynchronous commands}\label{sec:time_delayed_correction}

From phase estimates $\widehat{\phivec}_{k|k}(\alphavec)$ any future
disturbance can be computed using the predictive models above for
either the synchronous and asynchronous correction cases. The latter
is dealt with straightforwardly since now the commands are computed
for time step $k + \Delta$, for fractional delays either integer or
fractional 
multiples  of the integration time $T_s$. We further assume that due to
discrete nature os signals and sequential operations $\tau = T_s
+ \delta$ leading to $\Delta = \tau/T_s = 1 + \delta/T_s$, i.e. one
frame delay plus pure system lag.

\begin{figure}[htpb]
	\begin{center}
          \includegraphics[width=1.0\textwidth]{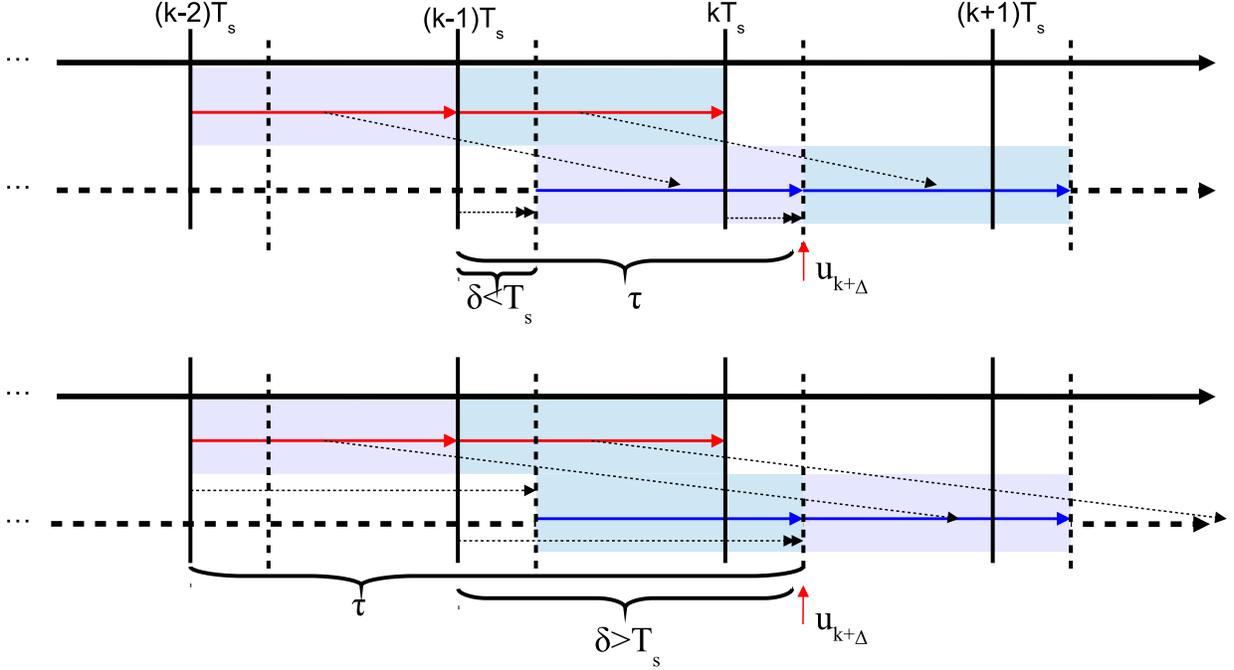}
	\end{center}
	\caption[]
	{\label{fig:temporal_diagram}
	Temporal diagrams. Top: $\delta < T_s$; Bottom: $\delta >
        T_s$. %The asynchronous commands are applied at $t=k T_s +\delta$  regardless of the value of $\delta$.% , providing $\Delta_n = \text{Int}(\tau/T_s)=0$, $\Delta_r \neq 0$. Bottom: $\tau > T_s$, with $\Delta_n=1$, $\Delta_r \neq 0$.  % Commands $\mathbf{u}_k$ are conditioned % a function of $u_k = f\left(\widehat\varphivec_{k+1}, \cdots, \varphivec_{k+1-p}\right)$ conditioned 
%         % to measurement $\svec_{k-1}$. Bottom: $T_s <\Delta < 2T_s$; $\uvec_k$ is conditioned to $\svec_{k-2}$.
% Also indicated are the weighted linear combinations of phase estimates if one where to use the consecutive state estimates directly \cite{poyneer08}.
      }
\end{figure}

 % $\Delta = \Delta_n + \Delta_r$ where $\Delta_n = Int(\tau/T_s)/T_s$ is the integer multiple of the sampling time and $\Delta_r = \tau - \Delta_n$ is the remainder such that $0 \leq \Delta_r \leq 1$.
 Both cases are shown in Fig. \ref{fig:temporal_diagram} providing the temporal sequence of operations. 

The controller is applied in real-time by computing, at iteration $k$ for either case $\delta \geq T_s$ or $\delta \leq T_s$,
\begin{subequations}\label{eq:single_rate_RToperations}
\begin{align}
 %\widehat{y}_{k|k-1}  & = C_v \widehat{x}_{k|k-1} \\
  \widehat{\phivec}_{k+1|k}(\alphavec) & = \Asa \widehat{\phivec}_{k|k-1}(\alphavec) + \Asa\mathcal{H}_\infty\left(\svec_k (\alphavec) -
     \D \widehat{\phivec}_{k|k-1}(\alphavec) \right) 
% \\
%   \widehat{\phivec}_{k+1|k} & = \mathcal{A}\widehat{\phivec}_{k|k} 
\\   
\uvec_{k+\delta/T_s}(\betavec_i) 
%& = \mathbf{F} \average{ \phivec(\betavec_i) \phivec(\alphavec)^\T }\average{  \phivec(\alphavec) \phivec(\alphavec)^\T }^{-1} \widehat{\phivec}_{ k+1+\Delta|k} \nonumber\\
%& = \mathbf{F} \CovMat_{\phivec_\betavec,\phivec_\alphavec}\CovMat_{\phivec_\alphavec,\phivec_\alphavec}^{-1} \mathcal{A}_\Delta\widehat{\phivec}(\alphavec)_{k|k} \nonumber \\
& = \mathbf{F} \CovMat_{\phivec_{\betavec_i},\phivec_\alphavec}\CovMat_{\phivec_\alphavec,\phivec_\alphavec}^{-1} 
 \Asa_{\delta}\widehat{\phivec}_{k+1|k}(\alphavec) \nonumber \\
& = \mathbf{T}_\delta \widehat{\phivec}_{k+1|k}(\alphavec)
\end{align}
\end{subequations}
where $\Asa_\delta$ estimates the lead wave-front $\delta$
seconds ahead in time to overcome any pure lag delay (3ms for the case
of Raven) and multiplication by the anisoplanatic filter
$\CovMat_{\phivec_\betavec,\phivec_\alphavec}\CovMat_{\phivec_\alphavec,\phivec_\alphavec}^{-1}=\average{
  \phivec(\betavec) \phivec(\alphavec)^\T }\average{
  \phivec(\alphavec) \phivec(\alphavec)^\T }^{-1}$ provides the
pupil-plane wave-front estimate in the science direction of
interest. This two-step scheme is equivalent to a single-step
prediction on account of the conditional expectation properties,
namely 
\begin{equation}
  E\{\Phi | S=s\} = E\{\Phi | E\{Y|S=s\}\} 
\end{equation}

The DM projection is made in parallel for all the $N_\beta$ science
directions by multiplying by the generalized inverse of the DM
influence matrix in Eq. \eqref{eq:CSI}.

%\section{SA-LQG setup and features}
%%%%%%%%%%%%%%%%%%%%%%%%%%%%%%%%%%%%%%%% 
% ------------section--------------------
%%%%%%%%%%%%%%%%%%%%%%%%%%%%%%%%%%%%%%%% 

\section{Static and predictive Spatio-Angular wave-front estimation}% with near-Markovian time-progression}
The state-space framework offers a suitable means to derive static
reconstructors against which we will compare the LQG formulation. The
presentation that follows is a brief review of the standard
minimisation criteria in AO that can be found elsewhere
\cite{piatrou07, whiteley98a, ellerbroek02, correia14}.  Here we
derive it seamlessly from choosing a specific combination of matrices
in the LQG controller to set the context for the simulation results
that ensue.

%%%%%%%%%%%%%%%%%%%%%%%%%%%%%%%%%%%%%%%% 
% ------------section--------------------
%%%%%%%%%%%%%%%%%%%%%%%%%%%%%%%%%%%%%%%% 
\subsection{Static case}

%\section{Strehl-optimal wave-front reconstruction and correction -- zonal representation}

It is a well established result that the static \textit{minimum mean square error} MMSE solution \cite{correia14}
\begin{align}\label{eq:hat_phi_Collapsed}
\widehat{\phivec}_\betavec & \triangleq
                             \CovMat_{\phivec_\betavec,\svec_\alphavec}\CovMat^{-1}_{\svec_\alphavec,
                             \svec_\alphavec} {\svec}_\alphavec \\
  & =  \average{ \phivec_\betavec \phivec_\alphavec^\T }
 \D^\T \left(\D \average{ \phivec_\alphavec \phivec_\alphavec^\T } \D^\T +
   \average{ \etavec \etavec^\T }\right)^{-1} {\svec}_\alphavec
\end{align}
can be derived from the LQG solution by taking $\mathcal{A}= \0$ and therefore the conditional expectation $\widehat{\phivec}_{k|k-1}=\0$ in Eq. (\ref{eq:single_rate_RToperations}). In such case the solution of the Riccati equation is $\Sigma_\infty = \Sigma_\varphivec$ and the Kalman gain boils down to 
\begin{equation}\mathcal{H'}_\infty = \average{ \phivec_\betavec \phivec_\alphavec^\T }
\D^\T \left(\D \average{ \phivec_\alphavec \phivec_\alphavec^\T } \D^\T +
  \average{ \etavec \etavec^\T }\right)^{-1}
\end{equation}
which provides the same resulting estimate phase vector as in Eq. (\ref{eq:hat_phi_Collapsed}).

\subsection{Predictive case}

To reduce the lag error, spatio-temporal prediction may be built-in the reconstructor. In such case one computes instead \cite{vogel04a}%\footnote{See \cite{vogel04a} on how the predictive step pops out of the open-loop reconstruction. }
\begin{align}\label{eq:spatioangular_hat_phi_predict}
\widehat{\phivec}_{k+\Delta}(\betavec) & \triangleq \average{ \phivec_{k+\Delta}(\betavec) \svec_{k}^\T(\alphavec) } \CovMat^{-1}_{\svec_\alphavec,
                             \svec_\alphavec}
% \average{ \phivec_{\betavec,k+\Delta} \phivec_{\alphavec,k}^\T }
% \D^\T + \nonumber
% \\ &   \hspace{30pt}\left(\D \average{ \phivec_{\alphavec,k} \phivec_{\alphavec,k}^\T } \D^\T \average{ \etavec_k \etavec_k^\T }\right)^{-1} {\svec}_{\alphavec,k}
\end{align}
where $\average{ \phivec_{k+\Delta}(\betavec) \svec_{k}^\T(\alphavec) }
= \average{ \phivec(\rhovec-\mathbf{v}\tau, \betavec)\svec^\T(\alphavec) }$ is the covariance matrix of pupil-plane phase between the science and guide-star directions at two time instants separated by a lag $\tau$. 

Equation (\ref{eq:spatioangular_hat_phi_predict}) can be split into a spatial estimate followed by a temporal prediction step, being written in the two-step form
\begin{align}
\widehat{\phivec}_{k+\Delta}(\betavec) & = \Asa_\Delta \widehat{\phivec}_{k}(\betavec) \\ & =
\Asa_\Delta\CovMat_{\phivec_\betavec,\svec_\alphavec}\CovMat^{-1}_{\svec_\alphavec,
                             \svec_\alphavec} {\svec}_\alphavec
% \\ & = \mathcal{A}\average{ \phivec_{\betavec,k} \phivec_{\alphavec,k}^\T }
% \D^\T \nonumber \\ & \hspace{30pt}\left(\D \average{ \phivec_{\alphavec} \phivec_{\alphavec}^\T } \D^\T +
%   \average{ \etavec \etavec^\T }\right)^{-1} {\svec}_\alphavec % \\
% & = \average{ \phivec_{\betavec,k+1} \phivec_{\alphavec,k}^\T }
% \D^\T \left(\D \average{ \phivec_{\alphavec} \phivec_{\alphavec}^\T } \D^\T +
%   \average{ \etavec \etavec^\T }\right)^{-1} {\svec}_\alphavec
\end{align}
Both this two-step approach or the all-at-once are equivalent as
pointed out before due to conditional expectation properties.
% as is shown in Appendix \ref{app:1step_vs_2step_prediction}.
It remains true however many steps are made as long as the starting
point is a pupil-plane phase estimate.% -- which differs slightly from
                                % the argumentation used before to
                                % point out the differences between
                                % the SA-LQG formulation with respect
                                % to the layered LQG.

\section{Computational aspects}
\subsection{Off-line computation of Spatio-Angular covariance matrices}\label{sec:CompCovMat}
The point-wise covariance matrix computation involves only spatial distances from which the spatio-angular covariances matrices are obtained analytically sampling the covariance function for von-K\'arm\'an turbulence % $C_\psi(\rho) = \average{\phivec \phivec^\T}$
\begin{equation}\label{eq:cov_fcn_zonal}
C_\psi(\rho) = \left(\frac{L_0}{r_0}\right)^{5/3}\times
\frac{\Gamma(11/6)}{2^{5/6}
  \pi^{8/3}}\left[\frac{24}{5}\Gamma\left(\frac{6}{5}\right)\right]\left(\frac{2
    \pi \rho}{L_0}\right)^{5/6}K_{5/6}\left(\frac{2 \pi \rho}{L_0}\right) 
\end{equation}
with $L_0$ the outer scale of turbulence, $r_0$ Fried's parameter,
$\Gamma$ the 'gamma' function and finally $K_{5/6}$ a modified Bessel
function of the third order. Parameter $\rho = |\boldsymbol{\rho}|$ is the distance between two
phase points in the bi-dimensional plane $\rho= \sqrt{\Delta x^2 + \Delta y^2}$.

For the multiple-layer case, the pupil-plane covariance functions are found from the weighted sum over the layers where the weighting is given by the relative strength of layer $l$ (also known as the fractional $r_0$).

\begin{figure}[htpb]
	\begin{center}
	% use packages: array            
        %  \begin{tabular}{cc}
            \includegraphics[width=1.0\textwidth]{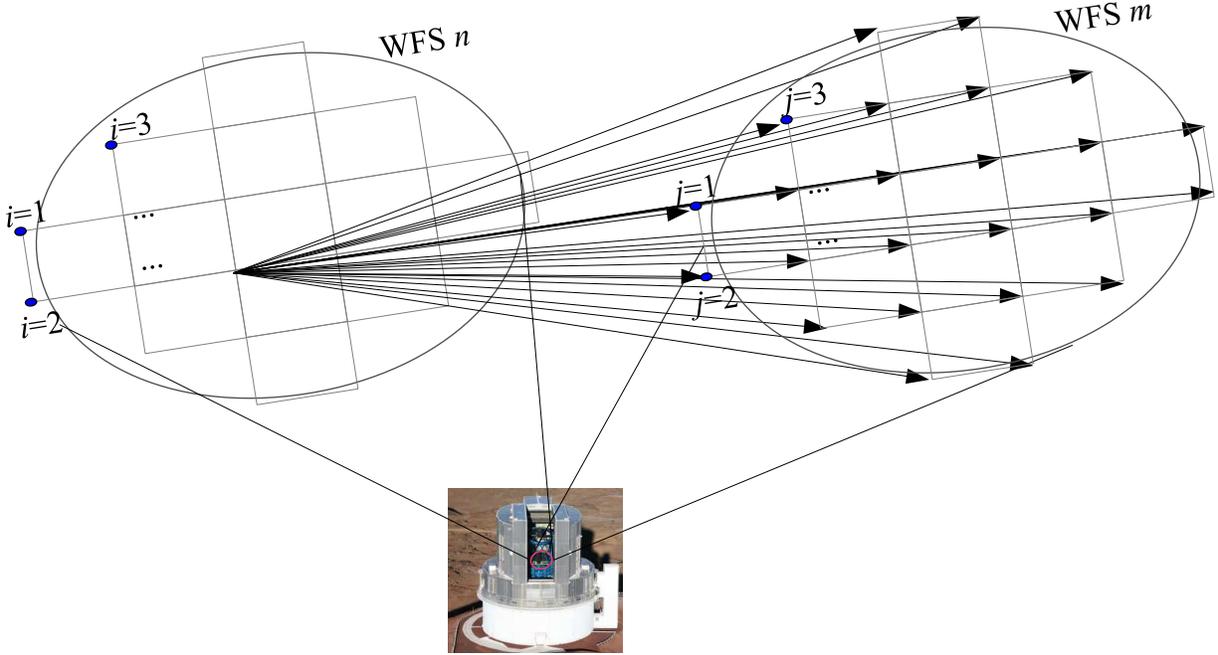}
         % \end{tabular}
	\end{center}
	\caption[]
	{\label{fig:SA_CovMat}
Spatio-angular computation scheme. Each column in the covariance matrix is obtained by sampling the phase covariance function with distances corresponding to the length of the arrows as shown. To comply with the 3x3 stencil in the gradient operator, the corners and the mid-points are sampled.}
\end{figure}

The spatio-angular covariance matrices are found by sampling and stacking the analytical covariance function in Eq. (\ref{eq:cov_fcn_zonal}) layer-by-layer (since layers are considered independent) between points tracing to the WFS sub-apertures' corners and mid-points whose coordinates 
in any layer are,$ [(\rho_{x_{i,n}} + \theta_{x_n} h_{\mathit{l}}),(\rho_{y_{i,n}} + \theta_{y_n} h_{\mathit{l}})$ 
and the separation vector between sub-aperture raster index $i$ on WFS $n$ and sub-aperture $j$ on WFS $m$ at layer $\mathit{l}$ is,
\begin{equation}
  \boldsymbol{\rho}_{i,j,l} = (\rhovec_{i,n} - \rhovec_{j,m} ) + h_{\mathit{l}} (\thetavec_n - \thetavec_m),
\end{equation}
where the first term of the right hand side of the equation is the separation between the sub-apertures in the pupil and the second term is the global WFS separation vector back-projected in the atmosphere. Figure \ref{fig:SA_CovMat} provides a schematic view of that procedure. 

%%%%%%%%%%%%%%%%%%%%%%%%%%%%%%%%%%%%%%%% 
% ------------section--------------------
\subsection{Real-time computational burden: scaling laws}
Regarding computational complexity, the SA-LQG formulation offers a
reduced number of real-time operations compared to the E-LQG
formulation. Furthermore it provides less than one order of magnitude
increase compared to MMSE (static or predictive) reconstructors for a ELT-sized case with a multiplexing factor of 20. 

% features an appealing scaling complexity with system size when compared to the static/predictive MMSE and the explicit LQG implementations. Figure \ref{fig:ratio_CompComplex_SALQG_vs_Others} shows the relative complexity for MOAO systems multiplexing up to 20 science directions. 

Take $\Ndm$ to be the number of DM or likewise science directions; $\Nact$ the number of valid actuators per DM; $\Nwfs$ the number of GS directions; $\Nslopes \approx 2\Nact$ the number of slope measurements per WFS; $\Nphi = 4\Nact$ is an under-estimate of the number of phase points estimated per reconstructed layer and $\Nl$ the total number of layers.
% \begin{itemize}
% \item MMSE: $(\Ndm \Nact) \times (\Nwfs\Nslopes)$
% \item LQG: $(\Nwfs\Nslopes)\times(\Nact\Nwfs) + (\Nwfs\Nact)\times(\Nwfs\Nslopes) + (\Nwfs\Nact)^2 + (\Ndm\Nact)\times(\Nwfs\Nact)$
% \item LQG Explicit: $(\Nwfs\Nslopes)\times(\Nact\Nl) + (\Nl\Nact)\times(\Nwfs\Nslopes) + \Nl(\Nact)^2 + (\Ndm\Nact)\times(\Nl*\Nact)$
% \item LQG Explicit Sparse: $(\Nwfs\Nslopes)\times(\Nact\Nwfs) + (\Nl\Nact)\times(\Nwfs\Nslopes) + \Nl(\Nact)^2 + (\Ndm\Nact)\times(\Nact)$
% \end{itemize}
Table \ref{tab:CompComplex} provides the approximate number of real-time multiply-and-accumulate (MAC) operations.
\begin{table}[h!]
\caption
{Approximate real-time computational complexity.}
\vskip 2mm
\begin{center}
\begin{tabular}{|l|c|c|c|}
\hline \hline
{\bf Op. / Algo.} & {\bf SA-LQG} & {\bf E-LQG} & {\bf MMSE}\\
\hline
$\mathcal{C}\widehat{\xvec}_{k|k-1}$ & $9 \Nslopes \Nwfs$ & $9 \Nslopes \Nwfs + 4 \Nphi \Nl$ & \\
$\Asa\mathcal{H}_\infty$  & $\Nwfs \Nphi \Nslopes \Nwfs$ & $\Nl \Nphi  \Nslopes \Nwfs$ & $\Ndm \Nact \Nwfs \Nslopes$ \\
$\Asa\xvec_{k|k}$ & $4 \Nphi \Nwfs$ & $4 \Nphi \Nl$ & \\
$\mathbf{F}\xvec_{k+1|k}$ & $\Ndm \Nact \Nphi$ & $\Ndm \Nact \Nphi + 4 \Nl \Nphi$
\end{tabular}
\end{center}
\label{tab:CompComplex}
\end{table}
If we take the case of Raven with $\Ndm=2$, $\Nact = 97$, $\Nwfs = 3$,
$\Nslopes = 144$ we get an increased computational complexity of about
a factor 15 for the SA-LQG and 30 for the E-LQG controller with 9
layers. 

\begin{figure}[htpb]
  \begin{center}
\includegraphics[width=1.0\textwidth]{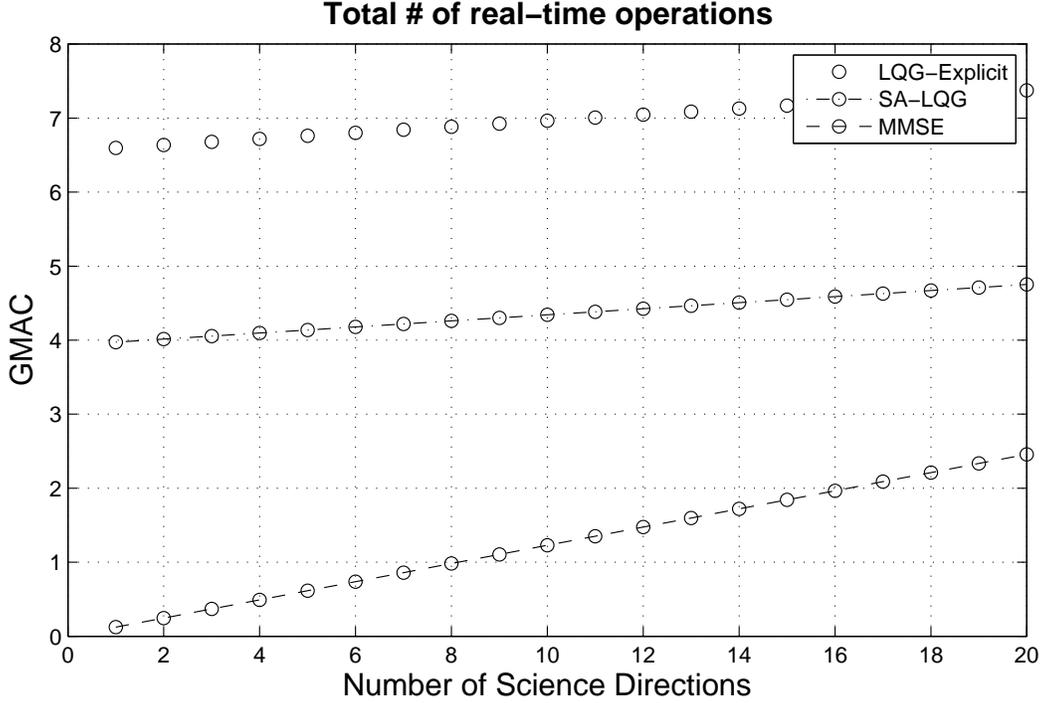} 
  \end{center}
  \caption[]
  {\label{fig:ratio_CompComplex_SALQG_vs_Others}
    Number of Giga
    multiply-and-accumulate (GMAC) real-time operations required by the LQG
    when compared to the static/predictive MMSE and the Explicit
    LQG. 3200 actuators, 6 WFS with 2x3200 slopes per WFS, phase estimated over 2x number of actuators per DM, 10 layers}
\end{figure}

As an example Fig. \ref{fig:ratio_CompComplex_SALQG_vs_Others} shows
the scaling complexity for ELT-sized systems (e.g. E-ELT MOSAIC \cite{14}, TMT's
IRMOS \cite{andersen06}) as a function of the number of science targets for $\Nact= 3200$ (from a 64x64 DM), $\Nslopes = 2\Nact$, $\Nphi = 4\Nact, $$\Nl = 10$ and $\Nwfs = 6$. For a maximum multiplex capacity of 20 science targets both the SA-LQG and the E-LQG are only a factor 2 and 3 more demanding than the MMSE, \textit{i.e.} roughly the same complexity. We have only considered the MMSE implemented as a full VMM since there is growing evidence that this option maps well into GPUs and should therefore be preferred to iterative implementations \cite{veran14}. For the time being we exclude from this analysis other considerations such as memory bandwidth and pipelinability.   

%%%%%%%%%%%%%%%%%%%%%%%%%%%%%%%%%%%%%%%% 
% ------------section--------------------
%%%%%%%%%%%%%%%%%%%%%%%%%%%%%%%%%%%%%%%% 
\section{Sample numerical models and simulations}
%{\color{red} We need a detailed description of the simulations to be done and in which way they differ from the ones in the previous paper. One should eventually fully use a Raven configuration and make a thorough assessment.

An end-to-end Monte Carlo wave-optics simulation of the Raven tomographic system in a zonal basis has been set up using the object-oriented OOMAO simulator \cite{conanr14} to examine the performance of the SA-LQG algorithm in relation to simpler static and predictive reconstructors.

% more basic algorithms in the same basis. These simulations take the work done in Zernike modal space in \cite{correia14} one step further to implement the full open-loop LQG algorithm for MOAO, comparing the Spatio-Angular method presented here to simpler static and predictive methods in simulation and in laboratory tests.

% System modelling of Raven was carried out in two separate parts. The first part involved the study of a broad swath of parameter space using two independent simulation tools. The parameter space was explored in order to establish and/or verify design parameters, as well as determine if Raven can realistically meet the proposed performance requirements and deliver useful MOAO-corrected images to the Subaru IRCS spectrograph\cite{andersen_jackson12}. The second part focussed specifically on implementing tomographic reconstructor algorithms with the intention of improving on the baseline performance case. 

%The two independent simulation platforms used to undertake the analysis were: MAOS (Multi-threaded Adaptive Optics Simulator), an AO simulation tool developed in C by Lianqi Wang and TMT, and OOMAO (Object Oriented MATLAB Adaptive Optics), a MATLAB object-oriented AO simulation tool developed by Rodolphe Conan at the UVic AO Lab with additional contributions from Kate Jackson, Peter Hampton and Olivier Lardiere. For a more detailed description of the MAOS code and the results of the parameter space study, see \cite{andersen_jackson12}.

\subsection{Model Parameters and Validation}

The figure-of-merit we use to establish the quality of the wavefront correction is the ensquared energy (EE); this is computed as the ratio of the amount of light falling on a science camera within a given angle projected on-sky to the total amount of light reaching the detector. Strehl ratios are also computed for completeness.

Table \ref{tab:baseline} contains the main simulation parameters, selected to reflect the physical properties available in the Calibration Unit of our Raven system test-bench \cite{andersen12a}. These include a three layer atmosphere located at ground, 5.5 and 11\,km altitudes, an turbulence coherence length $r_0$ of 19\,cm and an atmospheric outer scale $L_0$ of 40\,m. 

\begin{table}[h!]
\caption
{Raven Baseline Configuration Parameters.}
\vskip 2mm
\begin{center}
\begin{tabular}{ll}
\hline \hline
{\bf Telescope} & \\
D & 8 m \\
\hline
{\bf Atmosphere} & \\
$r_0$ &  19\,cm \\
$L_0$ & 40 m \\
zenith angle & 0 deg \\
Fractional $r_0$ & [0.596; 0.224; 0.180]\\
Altitudes & [0, 5.5, 11]\,km\\
wind speeds & [5.68; 6; 17]\,m/s \\
wind direction & [90; 180;180]\,deg \\
\hline
{\bf Wavefront Sensor} & \\
% MOAO Mode & Enabled \\
$N_{NGS}$ & 3 \\
Order & 10$\times$10 \\
$\theta_{pix}$ & 0.4 arcsec \\
RON & 0.2\,e$^-$ \\
NGS radii & 45\,arcsec \\
$N_{pix}$ & 12 \\
$f_{sample}$ & 30-200\,Hz \\
$\lambda_{WFS}$ & 0.64\,$\mu$m \\
Centroiding algorithm & thresholded Centre-of-Gravity\\
\hline
{\bf DM} & \\
$N_{DM}$ & 2 \\
Order & 11$\times$11 \\
stroke & infinite \\
influence & cubic \\
\hline
{\bf AO loop} & \\
pure delay  & $\tau_\mathsf{lag}=$\,3ms \\
\hline
{\bf Evaluation Wavelength} & \\
%$N_{points}$ & 49 \\
$\lambda_{evl}$ & 1.65\,$\mu$m \\
%Sampling$_{PSF}$ & $\lambda/4/D$ \\
\hline
\end{tabular}
\end{center}
\label{tab:baseline}
\end{table}

Model behavior was verified by collecting a simulated long exposure science image using no AO correction mode. The theoretical estimate of the Full Width at Half Max (FWHM) of the long exposure PSF for the baseline atmosphere in J band (1.2 $\mu m$) can be computed for a finite outer scale using the expression developed in \cite{tokovinin02} for the ratio, $L_0/r_0 > 20$. With $r_0 = 0.156cm$ at $0.5 \mu m$ and $L_0 = 30 m$ (chosen from median conditions on Mauna Kea), this computation yields an expected FWHM of 0.3891 arc seconds in J-band. The FWHM of a simulated long exposure science image was computed to be 0.3865 arc seconds, confirming good agreement between theory and simulation. Note that this test was carried during initial validation of the simulation tools before parameters such as $r_0$ were finalized.
%\ccc{Table parameters are for ro=19cm, whereas FWHM result is for r0=15.6cm. Inconsistent?}

\subsection{Results}

Previous work \cite{correia14} examined the trade-off between lower WFS frame-rates which reduces spatial errors due to noise propagation since  measurement's SNR is higher but increases temporal lag error. Correction quality was shown to increase with WFS integration time to a point and then begin to decrease as the lag error exceeded the SNR gain. 

Table \ref{tab:numericalResultsZonalPred} summarizes the peak performances over the frame-rates for varying GS magnitudes. 
% The zonal static and predictive reconstructors discussed in Sec. \ref{sec:StaticPredRecs} were implemented in simulation and the framerate at-which the peak performance was achieved for each reconstructor at increasing GS magnitudes was determined; the results of the simulations are summarized in Table. \ref{tab:numericalResultsZonalPred}. 

\begin{table}[h!]
\caption
{Raven End-to-End simulation results. The optimal performance ( \% ensquared energy) for each GS magnitude is shown for the zonal static SA, and compared to stand-alone SA Prediction and the full SA LQG algorithm.}
\vskip 2mm
\centering
\begin{tabular}{ l|cccc|cccc|cccc}
\hline
GS mags & \multicolumn{4}{c|}{static SA} & \multicolumn{4}{c|}{SA Prediction} & \multicolumn{4}{c}{SA LQG} \\

 & EE & lag & Strehl & lag & EE & lag & Strehl & lag & EE & lag & Strehl & lag \\
\hline
\hline
13.5 & 47.99 & 6 & 28.55 & 13 & 49.88 & 23 & 31.70 & 15 & 52.27 & 20 & 33.54 & 15 \\
14 & 47.48 & 10 & 28.05 &14 & 50.01 & 25 & 32.32 & 22 & 52.66 & 22 & 34.60 & 22 \\
14.5 & 46.79 & 12 & 27.98 & 15 & 49.65 & 24 & 31.72 & 24 & 52.48 & 24 & 34.02 & 23 \\
15 & 45.73 & 13 & 25.64 & 15 & 49.07 & 25 & 30.66 & 25 & 52.06 & 25 & 33.12 & 25 \\
15.5 & 44.36 & 15 & 23.41 & 18 & 48.41 & 26 & 29.11 & 25 & 51.40 & 25 & 31.53 & 25 \\
16 & 42.57 & 16 & 21.52 & 22 & 47.31 & 30 & 26.50 & 25 & 50.27 & 26 & 29.25 & 27 \\
16.5 & 40.43 & 20 & 18.90 & 25 & 46.73 & 41 & 24.33 & 40 & 48.98 & 35 & 26.72 & 35 \\
17 & 38.28 & 25 & 15.57 & 25 & 45.20 & 46 & 23.29 & 47 & 47.48 & 45 & 24.95 & 45 
\label{tab:numericalResultsZonalPred}
\end{tabular}
\end{table}

\begin{figure}[htpb]
  \begin{center}
    \includegraphics[width = 0.8\textwidth]{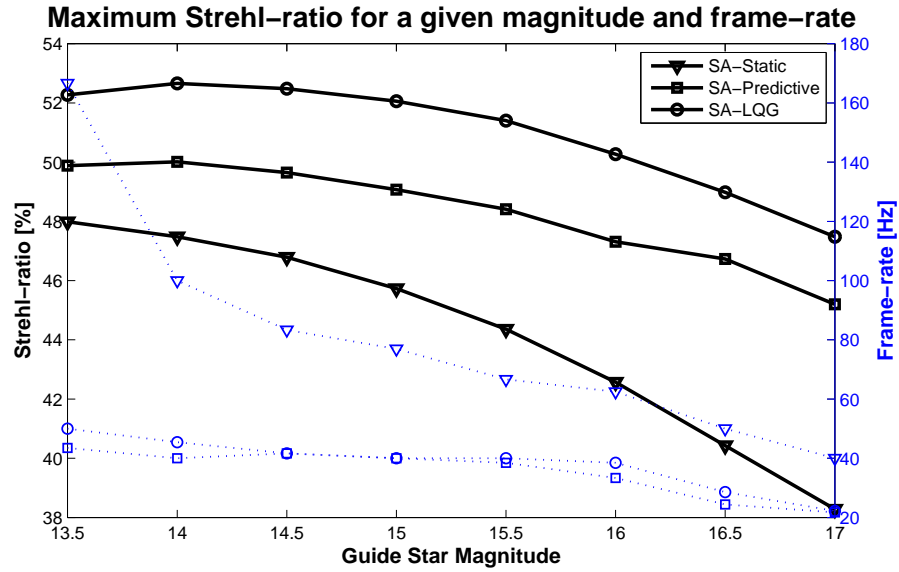} 
    \includegraphics[width = 0.8\textwidth]{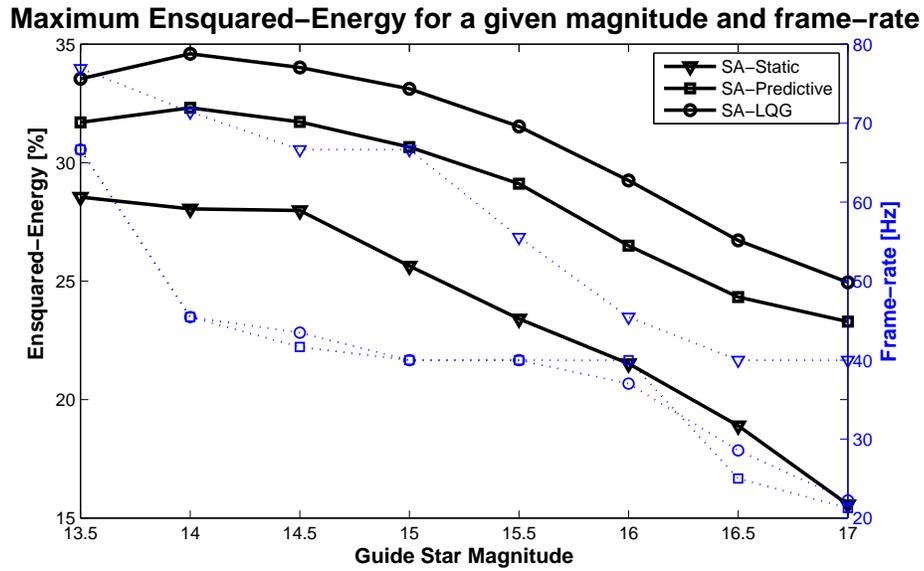}
  \end{center}
  \caption[]
  {\label{fig:LQG_Pred_static}
    Static reconstructor vs Stand-alone Prediction and LQG: Strehl ratios (top) and ensquared energy (bottom) as a function of NGS magnitude. Left axis shows best achieved performance with a static reconstructor compared to the predictive and LQG reconstructors; right axis shows the OL-WFS sample rate at which that performance is obtained.}
\end{figure}

These tests show that, as in the modal case, better peak performance
can be achieved using temporal prediction and this peak occurs at
lower frame-rates than the static algorithm for each
magnitude. Providing 
reduced noise-propagation the SA-LQG algorithm shows that the best performance occurs at approximately the same WFS frame-rate as in the predictive case, but that overall performance is significantly improved, indicating a reduction in spatial error. The data is also plotted in Fig. \ref{fig:LQG_Pred_static}; here it can be seen that by using temporal prediction equivalent performance can be achieved with higher magnitude GSs when using a lower frame-rate. The LQG can deliver performance equivalent to that of the predictive algorithm using GSs which are one magnitude dimmer and equivalent to the static algorithm using GSs which are greater than two magnitudes dimmer. 

% \ccc{In passing, let us note that the theoretical estimate of the Full Width Half Max (FWHM) of the long exposure PSF for the baseline atmosphere in J band (1.2 $\mu m$) can be computed for a finite outer scale using the expression developed in \cite{tokovinin02} for the ratio, $L_0/r_0 > 20$. With $r_0 = 0.156cm$ at $0.5 \mu m$ and $L_0 = 30 m$ (chosen from median conditions on Mauna Kea), this computation yields an expected FWHM of 0.3891 arc seconds in J-band. The FWHM of a simulated long exposure science image was computed to be 0.3865 arc seconds, confirming good agreement between theory and simulation.}

%The results show that the framerate at-which peak performance occurs for both the predictive and LQG algorithms is slower than that of the static algorithm for each magnitude; not only that, but the overall performance at each GS magnitude is improved with the predictive algorithm and again with the LQG. The data is plotted in Fig. \ref{fig:zonalPred_Results}; here it can be seen that by using temporal prediction equivalent performance can be achieved with higher magnitude GSs when using a lower framerate.\ccc{ Is seems to be 2 and higher magnitudes gain with the LQG, is this so? I believe we should state it here (and also in the abstract.}

\subsection{SA-LQG v. E-LQG}\label{sec:SALQG_vs_ELQG}

In appendix \ref{app:SA_vs_Explicit_LQG}  it is shown that the SA-LQG
can be deduced from the E-LQG (under certain conditions) in a stationary regime. 

The geometrical nature of the problem, however, is such that we can
retrieve the pupil-plane wave-front from its layered counterpart
through ray-tracing but not the opposite. This is patent in the
noninvertibility of $\Proj$ which imposes a spatial truncation on the
wave-fronts over the
footprints in the GS direction otherwise represented on larger
meta-pupils. 

% Monte-Carlo simulations do indeed show small discrepancies that lead to a
% sub-percent difference in both SR and EE when comparing performance in
% the GS directions ($\alphavec$). 

% We believe that such differences are in relation with the
% near-Markovianity of the model for reasons similar to the ones pointed
% out in \S \ref{sec:nearMarkovModel}, but a complete understanding will
% be left for a subsequent analytical analysis. The same effects are
% revealed in the invertibility of the ray-tracing operators. Although
% we found several cases where we could retrieve correct results, in
% general the invertibility of $\Proj$ is asterism-dependent and not
% assured.

Simulations have shown that
\begin{equation}
%EE(\psi_\alphavec, \widehat{\psi}_{k|k}(\alphavec)) = EE(\psi_\alphavec, \Proj_\alphavec \widehat{\varphivec}_{k|k})
\average{\Vert \psi_k (\alphavec)  - \widehat{\psi}_{k|k}(\alphavec)
  \Vert^2} \geq \average{\Vert \psi_k (\alphavec)  - \Proj_\alphavec \widehat{\varphivec}_{k|k}\Vert^2}
\end{equation}
i.e., the averaged residual phase variance in the GS directions obtained with the SA-LQG is
higher than the one obtained with the E-LQG. This difference then
translates to the science when using the anisoplanatic filter as
follows
\begin{equation}
\average{\Vert \psi_k (\betavec)  - \CovMat_{\rhovec_\betavec
    ,\rhovec_\alphavec}\CovMat_{\rhovec_\alphavec,\rhovec_\alphavec}^{-1}
  \widehat{\psi}_{k|k}(\alphavec)
  \Vert^2} \geq 
 \average{\Vert \psi_k (\betavec)  - \Proj_\betavec \widehat{\varphivec}_{k|k}\Vert^2}
%EE(\psi_\alphavec, \widehat{\psi}_{k|k}(\alphavec)) = EE(\psi_\alphavec, \Proj_\alphavec \widehat{\varphivec}_{k|k})
\end{equation}
The anisoplanatic filter \textit{per se} adds no further error to the
estimation since 
\begin{equation}
\average{\Vert \psi_k (\betavec)  - \CovMat_{\rhovec_\betavec
    ,\rhovec_\alphavec}\CovMat_{\rhovec_\alphavec,\rhovec_\alphavec}^{-1}
  \Proj_\alphavec \widehat{\varphivec}_{k|k}
  \Vert^2} = 
 \average{\Vert \psi_k (\betavec)  - \Proj_\betavec \widehat{\varphivec}_{k|k}\Vert^2}
%EE(\psi_\alphavec, \widehat{\psi}_{k|k}(\alphavec)) = EE(\psi_\alphavec, \Proj_\alphavec \widehat{\varphivec}_{k|k})
\end{equation}
 
A modal decomposition of the errors using Zernike polynomials showed
that the low-order modes such as tilts, focus and astigmatism are more
affected. This is also in agreement with the little impact in EE
observed which would suffer more from poorer correction of high order modes.

For the Raven case, the differences observed in the science direction are of the percent level.

\section{Raven Laboratory Test Results}

%Science images in J-Band were taken for a wide asterism case using both GLAO and MOAO; SCAO and no AO cases are included for comparison.
%
%\begin{figure}[h!]
%\centering
%\includegraphics[width = 1.2\textwidth]{scienceImsFilter3.png}
%\caption{J-Band Science Images of GLAO vs MOAO for a static zonal SA tomographic reconstructor. Asterism diameter was 2 arc min and NGS magnitudes were 11.25,12.13, and 12.13.}
%\label{fig:staticSciImgs}
%\end{figure}

\begin{figure}
\centering
\includegraphics[width = 1.0\textwidth]{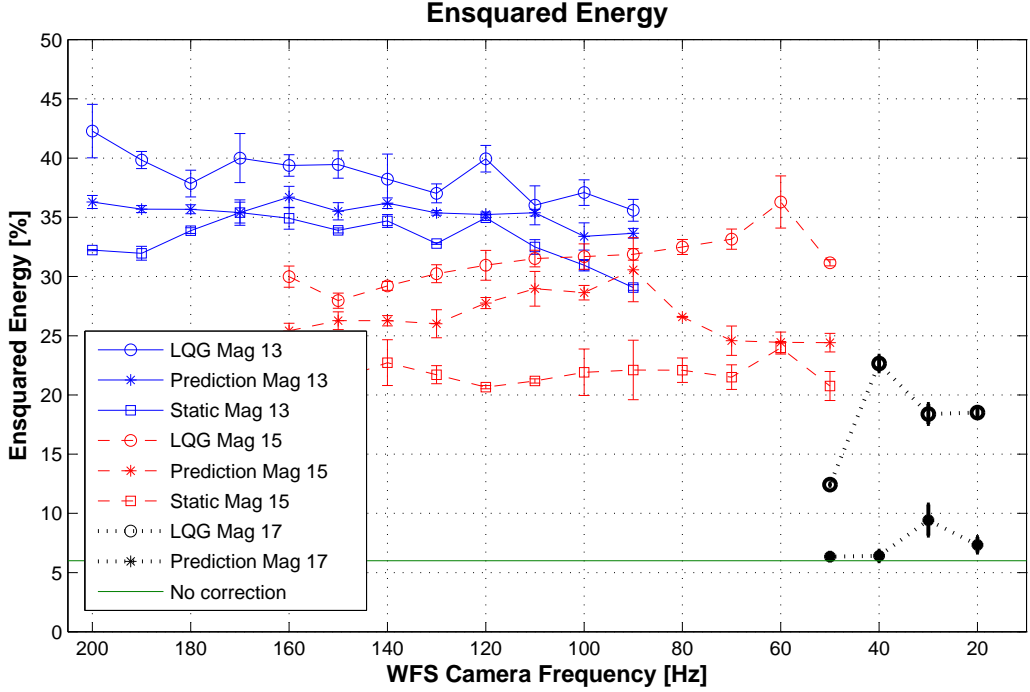}
\caption{Measured ensquared energy in a 140 milliarcsecond slit of
  science images taken while running the NGS WFSs at varying frame
  rates and executing static, predictive and LQG algorithms at three
  different magnitude settings. The static reconstructor results are not shown
for the faintest stars as the performance is not better than no AO
correction.}
\label{fig:recCompare}
\end{figure}

Raven's physical design and system capabilities are summarized in
\cite{jackson12, andersen12a, andersen12}; it is equipped with a
calibration unit (CU) which also serves as a telescope simulator and
turbulence generator. A broad spectrum source illuminates an array of
pinholes, any three of which can be picked off to create an NGS
asterism, and a series of filters which allow incremental changes in
the magnitudes of all asterism stars at once. A cross section of
measurements across NGS WFS frame rates were taken for three different
filters to show the improvement in performance using prediction and
SA-LQG control over the static reconstructor. The laboratory results
also validate our simulation results which indicate that, using the
SA-LQG reconstructor, the system can produce performance equivalent to
that obtained using the static reconstructor when GSs are 2 magnitudes
fainter. A plot of the results of these measurements is shown in
Fig. \ref{fig:recCompare}. The diameter of the asterism used is
approximately 2 arminutes. The CU lamp does not illuminate all
pinholes with uniform intensity, therefore the data points labeled Mag
13 were generated using GSs with magnitudes [12.6,13.25, 13.25]; those
labeled Mag 15 are actually [14.3, 15, 15], and those labeled Mag 17
are [16.5, 17.2, 17.2].

%\ccc{More comments here. E.g. For the magnitude 13 asterism the peak performance is at ~70\,Hz frame-rate for the static whereas it is 30\,Hz for the predictive algorithm and 60 for the LQG. This is "curious" because it shows we can go slower to increase performance with the predictive but the LQG can do (and actually increase perf) at a higher frame-rate. For the dim asterism case the trend is not the same but there is a 11\% EE increase with the LQG at 20\,Hz!!. Also, all results stated at the 1.65micron wave-length?}

The performance achieved in lab tests is lower than that predicted by simulations in all reconstructor cases. However the trend of increasing peak performance with each increase in reconstructor complexity is reflected in both the simulation data and the measured data. The overall decrease in performance can be attributed to multiple sources; these include imperfect calibration, which has a significant impact on OL systems, underestimation of noise sources in the simulation compared to reality, effects of the rotated WFSs, DM fitting, OL go-to errors, and non-common path aberrations (NCPA) between the OL and science paths, as well as between the CL-WFSs and the science camera. A higher amount of total system lag than anticipated may also be reducing the performance of all three types of reconstructors. Albeit, the main simulation findings in improved performance could be reproduced with the test-bench. We are currently investigating ways of enhancing the calibration and fine knowledge of system parameters in order to optimize overall system performance and gain understanding over unmodelled system features.

%A trend noted in the laboratory measurements for brighter GSs is that the peak performance of the static and LQG reconstructors occurs at rates slightly slower than the frequencies predicted by simulation, but the simple predictor performance peaks at a much slower frequency than expected. The likely cause of this trend is greater noise in the laboratory test setting which  would be expected to shift the peak performance of all reconstructors both down in value and to lower frequency, but to have the most effect on the stand-alone predictor since it only mitigates temporal lag error while the LQG mitigates both noise and lag. \ccc{I have a back of the envelope calculation which shows that the lag time at which the best performance is achieved for the predictor increases more quickly with increased noise variance than the other two. We can discuss tomorrow - I will type it up in a note.} 

\begin{figure} [h!]
\centering
\includegraphics[width = 1.0\textwidth]{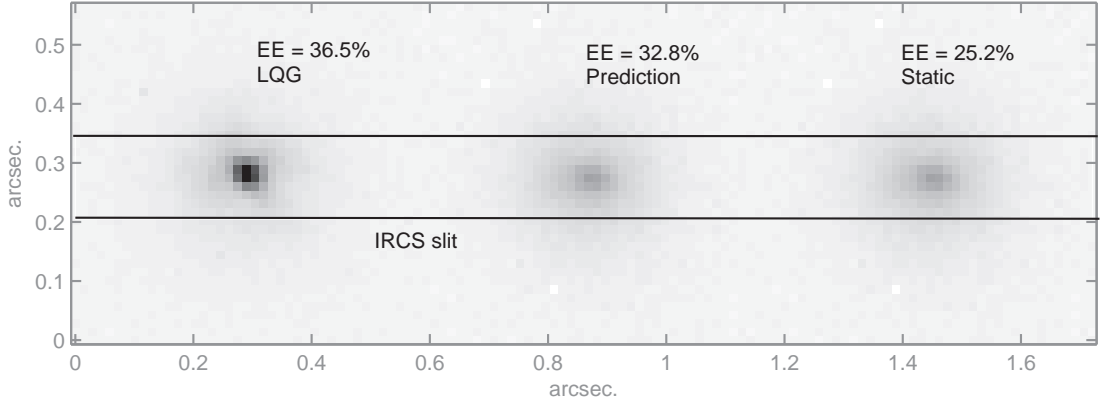}
\caption{Comparison of Science camera images for SA-LQG, Predictive
  and Static tomographic reconstruction algorithms. Peak frequency PSF shown for each reconstructor using magnitude 15 NGSs (see Fig. \ref{fig:recCompare}) .}
\label{fig:labSciImgs}
\end{figure}

The science images in Fig. \ref{fig:labSciImgs} represent the best performance achieved for each reconstruction method for a fixed magnitude; they clearly show that the EE is increased and the spot image becomes smaller for both predictive and LQG algorithms over the static reconstructor.

% \begin{figure}[h!]
% \centering
% \includegraphics[width = 0.5\textwidth]{bestSciImsFilter4EE.ps}
% \caption{Performance comparison for Static, Predictive and LQG tomographic reconstruction algorithms.}
% \label{fig:labSciImgs}
% \end{figure}

%The performance achieved in lab tests is lower than that predicted by simulations in all reconstructor cases, but the trend of increasing peak performance by between 2 and 3 \% with each increase in reconstructor complexity can be seen to be reflected in both the simulation data and the measured data. The exception is the static reconstructor at magnitude 15 in the laboratory results; at this point, the signal on the WFSs becomes quite low and the thresholded CoG begins to fail in the lab setting.

%The overall decrease in performance can be attributed to any number of sources; these include imperfect calibration, underestimation of noise sources in the simulation compared to reality, effects of the rotated WFSs, DM fitting, OL go-to errors, and NCPA between the OL and science paths, as well as between the CL WFSs and the science camera. An investigation into the effects of these errors on simulation results is under way.
%}

%%%%%%%%%%%%%%%%%%%%%%%%%%%%%%%%%%%%%%%% 
% ------------section--------------------
%%%%%%%%%%%%%%%%%%%%%%%%%%%%%%%%%%%%%%%% 
\section{Summary}
We have outlined the dynamical Strehl-optimal LQG controller providing minimum
residual phase variance tailored to MOAO systems. The latter does not require explicit knowledge of the tridimensional wave-front
profile allowing for a simplified forward model. Our implementation uses anisoplanatic filters to estimate
pupil-plane wave-front profiles from any other directions based on
knowledge of the turbulent $C_n^2$ profile. One such approach can also
be useful for laser tomography AO.

We have included temporal prediction by
assuming frozen-flow, in which case time evolution can be converted to
spatial translations. We have proposed a minimal space-state representation
with a single temporal instance providing reduced computational
complexity. 

By allowing the controller to run slower while performing temporal
prediction, we can increase the SNR in the wave-front measurements by
collecting more light in each sensor's sub-apertures. We have shown
that we could gain two magnitudes for a same performance obtained with
a MMSE static reconstructor, leading to a sky-coverage improvement of
roughly a factor 5.

%%%%%%%%%%%%%%%%%%%%%%%%%%%%%%%%%%%%%%%%%%%%%%%%%%%%%%%%%%%%%
\section*{Acknowledgements}		

The research leading to these results 
received the support of the A*MIDEX
project (no. ANR-11-IDEX-0001- 02) funded by the "Investissements
d’Avenir'' French Government program, managed by the French National
Research Agency (ANR). 
C. Correia acknowledges the support of the European Research Council through the Marie Curie Intra-European Fellowship with reference FP7-PEOPLE-2011-IEF, number 300162.
%The research leading to these results has received funding from the European Union Seventh Framework Programme
%(FP7/2007-2013)  through the Marie Curie Intra-European Fellowship under grant agreement no. [FP7-PEOPLE-2011-IEF-300162].

% The work leading to this invention has received funding from the People
% Programme (Marie Curie Actions) of the European Union's Seventh
% Framework Programme (FP7/2007-2013) under REA grant agreement n°
% [xxxxxx].10

All the simulations and analysis done with the object- 
  oriented MALTAB AO simulator (OOMAO) \cite{conanr14} freely available from \htmladdnormallink{https://github.com/rconan/OOMAO/}{https://github.com/rconan/OOMAO/}

%%%%%%%%%%%%%%%%%%%%%%% References %%%%%%%%%%%%%%%%%%%%%%%%%
\bibliographystyle{osajnl}   %>>>> makes bibtex use osajnl.bst
\bibliography{references}   %>>>> bibliography data in references.bib

\appendix
%%%%%%%%%%%%%%%%%%%%%%%%%%%%%%%%%%%%%%%% 
% ------------section--------------------
\section{Relationships between SA and Explicit LQG}\label{app:SA_vs_Explicit_LQG}
We set forth to show that the SA-LQG formulation provided in this paper shares important features with the  more standard ``explicit'' E-LQG formulation which involves the estimation of phase in the layers.

Let the E-LQG state update equation
\begin{align}
\widehat{\varphivec}_{k+1|k}& = \Aexp \widehat{\varphivec}_{k|k-1} + \Lexp (s_k - \D \Proj \widehat{\varphivec}_{k|k-1})
\end{align}
where upper/subscript '$\text{{E}}$' indicates matrices for the E-LQG case.

For the near-Markovian order 1 time-progression model assumed earlier, the driving noise statistics are such that 
$\CovMat_n^\mathsf{E} = \CovMat_\varphivec - \Aexp \CovMat_\varphivec \AexpT$. Multiplying out by the ray-tracing operator to the left and its transpose to the right equates to $\CovMat_n = \CovMat_\phivec - \Asa \CovMat_\phivec \Asa^\T$ by plausibly taking 
\begin{align}
%\CovMat_\infty^{SA} & \neq \Proj\CovMat_\infty \Proj^\T \\
%\mathcal{M}_\infty^{SA} & = \Proj \mathcal{M}_\infty \\%\text(DEMONSTRATION MISSING)\\ 
\Asa & = \Proj \Aexp \Proj^{\dag}
\end{align}
and assuming $\Proj^{\dag}$ to be computable and furthermore that
$\Proj\Proj^{\dag}=\I$. This assumption however may not hold since
$\Proj$ is non-squared and its structure gives rise to a rank
deficient matrix. We thus expect (and have evidence for) some differences using the SA-LQG with respect to the E-LQG formulation. %The exact extent and in favour to which formulation is not clear at first sight. % Monte Carlo simulations have shown ...

% which ensures the state noise covariance matrix holds in both the explicit and the SA case, i.e
% $\CovMat_n^\mathsf{LAY} = \CovMat_\varphivec^\mathsf{LAY} - \Aexp \CovMat_\varphivec^\mathsf{LAY} \AexpT$ and its SA counterpart $\CovMat_n = \CovMat_\varphivec - \Asa \CovMat_\varphivec \Asa^\T$ can be obtained from each other using the definitions above.

We now add the following
\begin{align}
\Proj \Aexp & = \Proj(\Proj^{\dag}\Asa\Proj)\\
& = \Asa\Proj
\end{align}
and
\begin{align}
\Proj \mathcal{L}_\infty & = \Proj \Aexp \Mexp\\
& = \Asa\Proj\Mexp \\
& = \Asa\mathcal{M}_\infty
\end{align}

Multiplying the state update equation on both sides by $\Proj$ to the left equates to
\begin{align}
\Proj \widehat{\varphivec}_{k+1|k}& = \Proj\Aexp \widehat{\varphivec}_{k|k-1} + \Proj\Lexp (\svec_k - \D \Proj \widehat{\varphivec}_{k|k-1})\\
& = \Proj (\Aexp -\Lexp\D\Proj)\widehat{\varphivec}_{k|k-1} + \Proj\Lexp \svec_k
\end{align}
With the above identities one gets $\Proj \Lexp \D\Proj\widehat{\varphivec}_{k|k-1} = \Asa \Msa \D\widehat{\phivec}_{k|k-1}$, yielding
\begin{align}
 \widehat{\phivec}_{k+1|k}& = \Asa (\I - \Msa\D) \widehat{\phivec}_{k|k-1} + \Msa  \svec_k
\end{align}
which is the SA formulation. 

We are left with the derivation of equivalence of Kalman gains and respective Riccati solutions, both computed off-line. The Riccati equation writes
\begin{align}
\Sigmaexp & = \Aexp \Sigmaexp \AexpT + \SigmaexpN - \Aexp \Sigmaexp \Proj^\T \D^\T \nonumber \\ & \left(\D \CovMat_\infty \D^\T + \CovMat_\eta\right)^{-1}
 \D \Proj \Sigmaexp  \Aexp^\T
\end{align}
Multiplying by $\Proj$ on the left and $\Proj^\T$ on the right one gets immediately 
\begin{align}
\SigmaSA & = \Asa \SigmaSA \Asa^\T + \SigmaSAN \\& - \Asa \SigmaSA  \D^\T \left(\D \SigmaSA \D^\T + \CovMat_\eta\right)^{-1}
 \D  \SigmaSA \Asa^\T
\end{align}
where we take $\SigmaSA = \Proj \Sigmaexp \Proj^\T$ with the state
noise covariance matrix in the SA case obtained from the E-LQG
using the definitions above. In practice this assumption does not hold in part due to the finite number of iterations used to compute the asymptotic Riccati estimation error covariance matrix $\CovMat_\infty$ and its E-LQG counterpart, in part since the geometry of the problem imposes a spatial truncation on baselines over which covariances are computed unfavourable to the SA-LQG.

The Kalman gain can then be obtained as
\begin{align}
\Proj \Mexp & = \Proj \Sigmaexp \Proj^\T \D^\T \left(\D \Proj \Sigmaexp \Proj^\T \D^\T + \CovMat_\eta\right)^{-1} \label{eq:SAKgain_fromELQG}\\
\Msa & = \SigmaSA\D^\T \left(\D \SigmaSA \D^\T + \CovMat_\eta\right)^{-1}
\end{align}
which is the straight SA formulation.% and (would) conclude the demonstration.

Further to the results shown in Sect. \ref{sec:SALQG_vs_ELQG}, by
using the Kalman gain for the SA-LQG from
Eq. \eqref{eq:SAKgain_fromELQG} does not change the performances
reported.   

\end{document}